\documentclass[twocolumn,aps,prb
]{revtex4-1}

\usepackage{times}
\usepackage{graphicx}
\usepackage{amsfonts}
\usepackage{amsmath, amsthm, amssymb}
\usepackage{dsfont}
\usepackage{color}
\usepackage{microtype}
\usepackage{tikz}
\usetikzlibrary{decorations.markings}
\usetikzlibrary{shapes,positioning,arrows}
\usepackage{bm}
\usepackage{siunitx}
\usepackage[colorlinks=true, allcolors=black, citecolor=blue, linkcolor=blue]{hyperref}

\renewcommand{\rm}[1]{\mathrm{#1}}
\newcommand{\su}{\uparrow}
\newcommand{\sd}{\downarrow}

\newcommand{\s}{\sigma}

\newcommand{\D}{\mathcal{D}}

\newcommand{\rmd}{\mathrm{d}}
\DeclareMathOperator{\Tr}{Tr}
\DeclareMathOperator{\tr}{tr}

\newcommand{\del}{\partial}

\newcommand{\vk}{{\bf{k}}}
\newcommand{\vq}{{\bf{q}}}
\renewcommand{\vr}{{\bf{r}}}
\newcommand{\vf}{v_\rm{F}}

\newcommand{\muti}{\mu_\rm{TI}}
\newcommand{\on}{\omega_n}
\newcommand{\On}{{\Omega_n}}

\renewcommand{\v}[1]{{\bf{#1}}}
\newcommand{\vs}{\bm{\s}}
\newcommand{\eps}{\epsilon}

\newcommand{\C}{\mathcal{C}}
\newcommand{\G}{\mathcal{G}}

\newcommand{\eg}{{e.g.}~}

\usepackage{soul}

\allowdisplaybreaks[1]

\begin{document}

\title{Inverse proximity effect in \texorpdfstring{$s$}{s}-wave and \texorpdfstring{$d$}{d}-wave superconductors coupled to topological insulators}

\author{Henning G. Hugdal}
\author{Morten Amundsen}
\author{Jacob Linder}
\author{Asle Sudb{\o}}
\email[Corresponding author: ]{asle.sudbo@ntnu.no}
\affiliation{Center for Quantum Spintronics, Department of Physics, NTNU, Norwegian University of Science and Technology, NO-7491 Trondheim, Norway}


\begin{abstract}
We study the inverse proximity effect in a bilayer consisting of a thin {$s$- or $d$-wave} superconductor (S) and a topological insulator (TI). Integrating out the topological fermions of the TI, we find that spin-orbit coupling is induced in the S, which leads to spin-triplet $p$-wave ($f$-wave) correlations in the anomalous Green's function {for an $s$-wave ($d$-wave) superconductor}. Solving the self-consistency equation for the superconducting order parameter, we {find that the inverse proximity effect can be strong for parameters for which the Fermi momenta of the S and TI coincide}. {The suppression of the gap is approximately proportional to $e^{-1/\lambda}$, where $\lambda$ is the dimensionless superconducting coupling constant. This is consistent with the fact that a higher $\lambda$} gives a more robust superconducting state. For an $s$-wave S, the interval of TI chemical potentials for which the suppression of the gap is strong is centered at $\muti = \pm\sqrt{2m\vf^2\mu}$, and increases quadratically with {the} hopping parameter $t$. Since the S chemical potential $\mu$ typically is high for conventional superconductors, the inverse proximity effect is negligible except for $t$ above a critical value. For sufficiently low $t$, however, the inverse proximity effect is negligible, in agreement with what 
has thus far been assumed in most works studying the proximity effect in S-TI structures. {In superconductors with low Fermi energies, such as high-$T_c$ cuprates with $d$-wave symmetry, we again find a suppression of the order parameter. However, since $\mu$ {is} much smaller in this case, a strong inverse proximity effect can occur at $\muti=0$ for much lower values of $t$. Moreover, the onset of a strong inverse proximity effect is preceded by an increase in the order parameter, allowing the gap to be tuned by several orders of magnitude by small variations in $\muti$.}
\end{abstract}

\maketitle

\section{Introduction}
Topological insulators are insulating in the bulk, but host metallic surface states protected by the topology of the material.\cite{Hasan2010,Qi2011,Wehling2014} For three-dimensional topological insulators, the two-dimensional (2D) surface states can be described by a massless analog of the relativistic Dirac equation, having linear dispersions and spin-momentum locking. Many interesting phenomena are predicted to occur by coupling the TI to a superconductor, thus inducing a superconducting gap in the TI.\cite{Alicea2012} For instance, such systems have been predicted to host Majorana bound states,\cite{Fu2008} which could be used for topological quantum computing. Moreover, the Dirac-like Hamiltonian $\vs\cdot\vk$ has consequences for the response to exchange fields, allowing the phase difference in a Josephson junction to be tuned by {an} in-plane magnetization to values other than $0$ and $\pi$,\cite{Tanaka2009} and inducing vortexes by an in-plane magnetic field.\cite{Zyuzin2016,Amundsen2018}

Numerous papers have studied the interesting phenomena {that have been discovered} in topological insulators with proximity-induced superconductivity.\cite{Fu2009,Akhmerov2009,Linder2010a,Linder2010b,Zhang2011,Cook2011,Qu2012,Cook2012,Sochnikov2013,Koren2013,Galletti2014,Sochnikov2015,Li2015,Kim2016} To our knowledge, however, much less attention has been paid  to the inverse superconducting, or topological,\cite{Shoman2015} proximity effect, i.e. the effect that the topological insulator has on the superconductor order parameter. There have been indications that superconductivity might be suppressed,\cite{Sochnikov2013} while other studies have found no suppression.\cite{Sochnikov2015} One recent study demonstrated that the proximity to the TI induces spin-orbit coupling in the S, possibly making a Fulde-Ferrel\cite{Fulde1964} superconducting state energetically more favorable near the interface of a magnetically doped TI.\cite{Park2017} Another study showed that the TI surface states can leak into the superconductor, resulting in a Dirac cone in the density of states.\cite{Sedlmayr2018} In this paper, we focus on the superconducting gap itself and study under what circumstances the inverse proximity effect is negligible, as is often assumed in theoretical works.

Using a field-theoretical approach, we study an atomically thin Bardeen-Cooper-Schrieffer (BCS) $s$-wave superconductor and $d$-wave superconductor coupled to a TI. While this is an approximation for most conventional and high-$T_c$ superconductors such as e.g. Nb, Al and YBa$_2$Cu$_3$O$_7$ (YBCO), superconductivity has been observed in e.g. single-layer NbSe$_2$\cite{Ugeda2016} and FeSe.\cite{Wang2012,Liu2012,He2013} Integrating out the TI fermions, we obtain an effective action for the S electrons. Due to the induced spin-orbit coupling, spin-triplet $p$-wave ($f$-wave) correlations are induced in the $s$-wave ($d$-wave) superconductor. 

Solving the mean-field equations, using parameters valid for both conventional $s$-wave superconductors and high-$T_c$ $d$-wave superconductors, we find that in both cases a strong suppression of the superconducting gap is possible. For conventional superconductors, where the Fermi energy $\mu$ is high compared to the cut-off frequency, the coupling between the S and the TI has to be quite large in order for the inverse proximity effect to be strong for relevant TI chemical potentials $\muti$. This can explain the lack of any inverse proximity effect in experiments.\cite{Sochnikov2015} {In high-$T_c$ $d$-wave superconductors, on the other hand, where the Fermi energy is much smaller, we find a strong gap suppression at much lower coupling strengths, which might therefore be experimentally observable.} For these systems, we also find an increase in the gap for $\muti$ just outside the region of strong inverse proximity effect.


The remainder of the article is organized as follows: The model system is presented in Sec.~\ref{sec:model}, and the effective action for the S fermions and order parameter is derived in Sec.~\ref{sec:eff_S}. In Sec.~\ref{sec:mean_field} we derive the mean field gap equations for the order parameter. Numerical results for the superconducting gap are presented and discussed in Sec.~\ref{sec:results}, and summarized in Sec.~\ref{sec:summary}. Further details on the calculation of the criteria for strong proximity effect, the Nambu space field integral, the zero-temperature, non-interacting gap solutions, {and the numerical methods used,} are presented in the Appendices.

\section{Model}\label{sec:model}
We model the bilayer consisting of a thin superconductor (S) coupled to a TI by the action
\begin{equation}
    S = S_\rm{S} + S_\rm{TI} + S_t.
\end{equation}
In {Matsubara and reciprocal space}, the superconductor is described by
\begin{align}
    S_\rm{S} ={}&\frac{1}{\beta V} \sum_k c^\dagger(k)\left(-i\on +\frac{\vk^2}{2m} -\mu\right)c(k) \nonumber\\
    &- \sum_{k,k',q} \frac{V_{\vk',\vk}}{(\beta V)^3}c_\su^\dagger(k') c_\sd^\dagger(-k'+q) c_\sd(-k+q) c_\su(k),
\end{align}
where $c(k) = [c_\su(k) ~ c_\sd(k)]^T$ with $c_{\su(\sd)}(k)$ denoting the annihilation operator for spin-up (spin-down) electrons,  $m$ is the electron mass, $\mu$ is the chemical potential in the S. {$\beta=1/k_B T$ and $V=L_xL_y$ are the inverse temperature and system area respectively. We have used the notation $k=(\on,\vk)$ ($q=(\On,\vq)$), where $\on$ ($\On$) is a fermionic (bosonic) Matsubara frequency, and $\vk$ ($\vq$) the fermionic (bosonic) in-plane wavevector.
$V_{\vk,\vk'}$ is the pairing potential, which can be written\cite{Fossheim2004}
\begin{equation}
  V_{\vk,\vk'} = g v(\vk)v(\vk'),
\end{equation}
where $v(\vk) = 1$ for $s$-wave pairing, and $v(\vk) = \sqrt{2}\cos(2\phi_\vk)$ for $d_{x^2-y^2}$-wave pairing, {where $\phi_{\vk}$ is the angle of $\vk$ relative to the $k_x$ axis}. The coupling constant $g$} is assumed to be non-zero only when $-\omega_-<\vk^2/2m - \mu<\omega_+$, where $\pm \omega_\pm$ is the upper (lower) cut-off frequency. For conventional $s$-wave superconductors this is typically taken to be the characteristic frequency $\omega_D$ of the phonons, while the cut-off frequencies in high-$T_c$ superconductors are of the order of the characteristic energy of the anti-ferromagnetic fluctuations {present in these materials}.\cite{Moriya1990,Monthoux1992,Pines1993,Moriya1994}
We will set $\hbar = 1$ throughout the paper. For the TI we use the Dirac action
\begin{equation}
    S_\rm{TI} = \frac{1}{\beta V}\sum_k\Psi^\dagger(k)(-i\on  + \vf\vk\cdot\vs - \muti)\Psi(k),
\end{equation}
where $\Psi(\vr) = [\psi_\su(\vr) ~ \psi_\sd(\vr)]^T$ describes the TI fermions, $\vf$ is the Fermi velocity, and $\muti$ is the TI chemical potential.
The S and TI layers are coupled by a hopping term\cite{BlackSchaffer2013,Takane2014,Park2017,Sedlmayr2018}
\begin{equation}
    S_t = -\frac{1}{\beta V}\sum_k t[c^\dagger(k)\Psi(k) + \Psi^\dagger(k)c(k)]. \label{eq:hopping}
\end{equation}
Similar models were recently used in Refs.~\onlinecite{Park2017,Sedlmayr2018} when studying similar systems with an $s$-wave S. {The full partition function of the system is therefore
\begin{align}
    Z = \int \D[c^\dagger, c] e^{-S_\rm{S}} \left(\int \D[\Psi^\dagger, \Psi] e^{-S_\rm{TI} - S_t}\right).
\end{align}}

\section{Effective action}\label{sec:eff_S}
As we are interested in the inverse proximity effect in the S and its consequences for the superconducting gap, we integrate out the TI fermions by performing the functional integral $Z_{\rm{TI},t} = \int \D[\Psi^\dagger,\Psi] e^{-S_{\rm{TI},t}}$,
where
\begin{align}
     S_{\rm{TI},t} ={}& \frac{1}{\beta V}{\sum_k} \Big\{\Psi^\dagger(k)(-G^{-1}_\rm{TI})\Psi(k)\nonumber\\
     &- t[c^\dagger(k)\Psi(k) + \Psi^\dagger(k)c(k)]\Big\}.
 \end{align}
Here, we have defined the matrix $G^{-1}_\rm{TI} = i\on - \vf\vk\cdot\vs+\muti$.
Performing the functional integration leads to an additional term in the S action,
 \begin{equation}
     \delta S_\rm{S} = \frac{t^2}{\beta V} \sum_k c^\dagger(k) G_\rm{TI} c(k),
 \end{equation}
 with the TI Green's function
 \begin{equation}
     G_\rm{TI} = \frac{i\on+\muti +\vf\vk\cdot\vs}{(i\on + \muti)^2-\vf^2\vk^2}.\label{eq:G_TI}
 \end{equation}
The effective S action thus reads
\begin{align}
    S_\rm{S}^\rm{eff} = {} &-\frac{1}{\beta V}\sum_k c^\dagger(k)G_0^{-1}c(k) \nonumber\\
    &- \sum_{k,k',q} \frac{V_{\vk',\vk}}{(\beta V)^3} c^\dagger_\su(k')c^\dagger_\sd(-k'+q) c_\sd(-k+q)c_\su(k),
\end{align}
where
we have defined the inverse non-interacting Green's function
\begin{equation}
    G_0^{-1} = i\on -\frac{\vk^2}{2m} + \mu - t^2G_\rm{TI}.
\end{equation}
From this we see that the coupling to $G_\rm{TI}$ in Eq.~(\ref{eq:G_TI}) leads to an induced spin-orbit coupling $\sim \vk\cdot\vs$ in the S, in agreement with Ref. \onlinecite{Park2017}.

Performing a Hubbard-Stratonovich decoupling,\cite{Altland2010} the 4-fermion term in the S action can be rewritten in terms of bosonic fields $\varphi(q)$ and $\varphi^\dagger(q)$,
\begin{align}
    &- \sum_{k,k',q}\frac{V_{\vk',\vk}}{(\beta V)^3} c^\dagger_\su(k')c^\dagger_\sd(-k'+q) c_\sd(-k+q)c_\su(k)\nonumber\\
     &\quad\rightarrow -\frac{1}{\beta V}\sum_{k,q}\left[\varphi(q)v(\vk)c_\su^\dagger(k)c_\sd^\dagger(-k+q) + \rm{h.c.}\right].
\end{align}
This also leads to an additional term in the total system action
\begin{equation}
    S_\varphi^0 = \frac{\beta V}{g}\sum_q\varphi^\dagger(q)\varphi(q),\label{eq:phi0action}
\end{equation}
and a functional integration of the bosonic fields in the partition function. Note that the decoupling is performed such that the bosonic fields have units of energy.

{By} defining the Nambu spinor
\begin{equation}
    \C(k) = [c_\su(k) ~ c_\sd(k) ~ c^\dagger_\su(-k) ~ c^\dagger_\sd(-k)]^T,\label{eq:nambu}
\end{equation}
the effective S action can be written
\begin{equation}
    S_\rm{S}^\rm{eff} = -\frac{1}{2\beta V} \sum_{k,k'} \C^\dagger(k) \G^{-1}(k,k') \C(k'),\label{eq:Seff_nambu}
\end{equation}
where
\begin{align}
    \G^{-1}(k,&k') =\nonumber\\
    & \begin{pmatrix}
        G_0^{-1}(k)\delta_{k,k'} & \varphi(k-k')v(\vk)i\s_y\\
        -\varphi^\dagger(-k+k')v(\vk)i\s_y & -[G_0^{-1}(-k)]^T\delta_{k,k'}
    \end{pmatrix}.\label{eq:Ginv}
\end{align}
Performing the functional integration over the fermionic fields, we arrive at the effective action for the bosonic fields
\begin{equation}
    S_\varphi = \frac{\beta V}{g} \sum_q \varphi^\dagger(q) \varphi(q) - \frac{1}{2}\Tr \ln (-\G^{-1}).
\end{equation}
The additional factor $1/2$ in front of the trace is due to the change in integration measure when changing to the Nambu spinor notation, see Appendix \ref{sec:Nambu_funcint} and \eg Ref.~\onlinecite{Krohg2018} for details.

\section{Mean field theory}\label{sec:mean_field}
Since $G_0^{-1}(i\on, \vk)$ is still inversion symmetric in the diagonal basis (see below), we assume that the bosonic field $\varphi(q)$ is temporally and spatially homogeneous as in the regular BCS case. However, a recent study has shown that introducing in-plane magnetic fields in the TI breaks this symmetry and can make a Fulde-Ferrel\cite{Fulde1964} order parameter energetically more favorable in an $s$-wave S.\cite{Park2017}
Calculating the matrix $\G(k)$ assuming a spatially homogeneous bosonic field $\phi(q) = \delta_{q,0}\Delta$, and defining the superconducting order parameter $\Delta(\vk) = \Delta \cdot v(\vk)$, we get
\begin{equation}
    \mathcal{G}(k) = \begin{pmatrix}
        G(k) & F(k)\\
        F^\dagger(k) & -G^T(-k)
    \end{pmatrix},
\end{equation}
where to leading order in $t$
\begin{widetext}
\begin{align}
    G(k) = {}& -\frac{\eps_\vk + i\on}{\xi_\vk^2+\on^2} - t^2\frac{(\eps_\vk+i \on)^2[(i\on+\muti)+\vf\vk\cdot\vs]}{(\xi_\vk^2+\on^2)^2[\vf^2\vk^2-(i\on+\muti)^2]} - t^2\frac{|\Delta(\vk)|^2[(i\on-\muti)-\vf\vk\cdot\vs]}{(\xi_\vk^2+\on^2)^2[\vf^2\vk^2-(i\on-\muti)^2]},\\
    F(k) ={}& \frac{\Delta(\vk)}{\xi_\vk^2 + \on^2}\bigg\{1+2t^2\frac{(\vf^2\vk^2-\muti^2-\on^2)\eps_\vk\muti - \on^2(\vf^2\vk^2+\muti^2+\on^2)}{(\xi_\vk^2 + \on^2)[(\vf|\vk|-\muti)^2+\on^2][(\vf|\vk|+\muti)^2+\on^2]}\nonumber\\
    &+2t^2\frac{(\vf^2\vk^2-\muti^2+\on^2)\eps_\vk - 2\on^2\muti}{(\xi_\vk^2 + \on^2)[(\vf|\vk|-\muti)^2+\on^2][(\vf|\vk|+\muti)^2+\on^2]}\vf\vk\cdot\vs\bigg\}i\s_y,\label{Fk}
\end{align}
\end{widetext}
with $\eps_\vk = \vk^2/2m - \mu$ and $\xi_\vk = \sqrt{\eps_\vk^2 + |\Delta(\vk)|^2}$. As mentioned above, the proximity-induced spin-orbit coupling leads to non-diagonal terms in $G(k)$. Moreover, $F(k)$ now has diagonal terms $\propto \vk\cdot\vs i\s_y$, signaling that {$p$-wave ($f$-wave) triplet superconducting correlations are induced in the $s$-wave ($d$-wave) superconductor. This has been shown to be the case in $s$-wave superconductors} when the spin-degeneracy is lifted by spin-orbit coupling.\cite{Gorkov2001} A similar expression was found for the anomalous Green's function on the TI side of an S-TI bilayer in Ref.~\onlinecite{Yokoyama2012}. {The results in Ref.~\onlinecite{Yokoyama2012} also suggest that odd-frequency triplet pairing could be induced in the S by including a magnetic exchange term $\v{m}\cdot\vs$ in the TI Lagrangian.}

\subsection{Gap equation}
While the above Green's functions contain information about the correlations in the superconductor, the superconducting gap must be determined self-consistently. We first change to the basis which diagonalizes the non-superconducting normal inverse Green's function $G_0^{-1}$. We find $G_{d,0}^{-1}(k) = P(k)G_{0}^{-1}(k)P^\dagger(k)$, where $G_{d,0}^{-1}(k) = \operatorname{diag}(G_{+,0}^{-1}(k), G_{-,0}^{-1}(k))$, with
\begin{equation}
    G_{\pm,0}^{-1}(k) = i\on -\eps_\vk - \frac{t^2}{i\on + \muti \mp \vf|\vk|}
\end{equation}
and
\begin{equation}
    P(k) = \frac{1}{\sqrt{2}}\begin{pmatrix}
        1 & e^{-i\phi_\vk}\\
        1 & -e^{-i\phi_\vk}
    \end{pmatrix}.
\end{equation}
Here $\phi_\vk$ is the angle of $\vk$ relative the $k_x$ axis. $+$ $(-)$ here denotes the Green's function for positive (negative) chirality states. Inverting $G_{d,0}^{-1}$ we find the Green's functions
\begin{equation}
    G_{\pm,0}(k) = \frac{i\on \mp\vf|\vk| +\muti}{[i\on -\eps_\pm^+(\vk)][i\on -\eps_\pm^-(\vk)]},
\end{equation}
where
\begin{align}
    \eps_\alpha^\gamma(\vk) ={}& \frac{1}{2}\big[\eps_\vk + \alpha \vf|\vk| - \muti \nonumber\\*
    &+ \gamma \sqrt{(\eps_\vk-\alpha\vf|\vk|+\muti)^2 + 4t^2}\big],\label{eq:nonsc_bands}
\end{align}
with $\alpha,\gamma = \pm 1$. The Green's function has residues
\begin{align}
    w_\alpha^\gamma(\vk) = \frac{1}{2}+\frac{\eps_\vk-\alpha \vf|\vk|+\muti}{2\gamma\sqrt{(\eps_\vk-\alpha\vf|\vk|+\muti)^2+4t^2}}.\label{eq:residues}
\end{align}

We next transform the entire inverse Green's function $\G$ using $\G_d^{-1}(k) = \mathcal{P}(k)\G^{-1}(k)\mathcal{P}^\dagger(k),$ where
\begin{equation}
    \mathcal{P}(k) = \begin{pmatrix}
        P(k) & 0 \\
        0 & P^*(-k)
    \end{pmatrix},
\end{equation}
which yields
\begin{equation}
    \G_d^{-1}(k) = \begin{pmatrix}
        G_{d,0}^{-1}(k) & -\Delta(\vk) e^{-i\phi_\vk}\s_z \\
        -\Delta^\dagger(\vk) e^{i\phi_\vk}\s_z & -G_{d,0}^{-1}(-k)
    \end{pmatrix}.
\end{equation}
Hence the full Green's function matrix for the superconductor is
\begin{equation}
    \G_d(k) = \begin{pmatrix}
        G_d(k) & F_d(k) \\
        F_d^\dagger(k) & -G_d(-k)
    \end{pmatrix},
\end{equation}
where we have defined the $2\times2$ matrices $G_d(k) = \operatorname{diag}(G_+(k), G_-(k))$ and $F_d(k) = \operatorname{diag}(F_+(k), F_-(k))$, and Green's functions
\begin{widetext}
\begin{subequations}
\begin{align}
    G_\pm(k) &= \frac{[i\on + \eps_\pm^+(\vk)][i\on + \eps_\pm^-(\vk)][i\on \mp \vf|\vk|+\muti]}{[i\on-\xi_\pm^+(\vk)][i\on+\xi_\pm^+(\vk)][i\on-\xi_\pm^-(\vk)][i\on+\xi_\pm^-(\vk)]}\label{Gdiag}\\
    F_\pm(k) &= \pm\frac{\Delta(\vk) e^{-i\phi_\vk}[(i\on)^2- (\pm \vf|\vk|-\muti)^2]}{[i\on-\xi_\pm^+(\vk)][i\on+\xi_\pm^+(\vk)][i\on-\xi_\pm^-(\vk)][i\on+\xi_\pm^-(\vk)]}.\label{Fdiag}
\end{align}
\end{subequations}
The eigenenergies of the system are now given by the poles in the above equation, where
\begin{align}
    \xi_\alpha^\gamma(\vk) = \frac{1}{\sqrt{2}}\Big\{\xi_\vk^2 + (\alpha \vf|\vk| - \muti)^2 + 2t^2 + \gamma \sqrt{[\xi_\vk^2 - (\alpha \vf|\vk| - \muti)^2]^2 + 4t^2[(\eps_\vk + \alpha \vf|\vk| - \muti)^2 + |\Delta(\vk)|^2]}\Big\}^{1/2}.
\end{align}
\end{widetext}

The gap equation for the amplitude $\Delta$ is found by requiring {$\frac{\delta S_\varphi}{\delta \Delta} = 0$,}\cite{Altland2010} which yields
\begin{equation}
    \Delta^\dagger =
    -\frac{g}{2\beta V}\sum_k \tr F_d^\dagger(k)v(\vk)\s_z e^{-i\phi_\vk}.
\end{equation}
Inserting the hermitian conjugate of Eq.~(\ref{Fdiag}) and performing the sum over Matsubara frequencies, we get the gap equation,
\begin{align}
    1 = {}& \frac{g}{4V} \sum_{\vk} v(\vk)^2\Big\{\frac{\xi_+^+(\vk)^2-(\vf|\vk|-\muti)^2}{\xi_+^+(\vk)[\xi_+^+(\vk)^2-\xi_+^-(\vk)^2]}\tanh\frac{\beta \xi_+^+(\vk)}{2}\nonumber\\
    &-\frac{\xi_+^-(\vk)^2-(\vf|\vk|-\muti)^2}{\xi_+^-(\vk)[\xi_+^+(\vk)^2-\xi_+^-(\vk)^2]}\tanh\frac{\beta \xi_+^-(\vk)}{2}\nonumber\\
    &+\frac{\xi_-^+(\vk)^2-(\vf|\vk|+\muti)^2}{\xi_-^+(\vk)[\xi_-^+(\vk)^2-\xi_-^-(\vk)^2]}\tanh\frac{\beta \xi_-^+(\vk)}{2}\nonumber\\
    &-\frac{\xi_-^-(\vk)^2-(\vf|\vk|+\muti)^2}{\xi_-^-(\vk)[\xi_-^+(\vk)^2-\xi_-^-(\vk)^2]}\tanh\frac{\beta \xi_-^-(\vk)}{2}
    \Big\}.\label{eq:gapeq}
\end{align}
Setting $t=0$ simply yields the regular BCS gap equation, which results in a gap $\Delta_0 = 2\omega_D e^{-1/\lambda}$ in the $s$-wave case,\cite{Bardeen1957} where $\lambda = g D_0/V$ is a dimensionless coupling constant, and $D_0$ is the density of states at the Fermi level.
$d$-wave pairing results in a slightly smaller gap for the same values for $\lambda$ and the cut-off frequencies, see Appendix \ref{sec:zerotempgap} for details. For $t\neq 0$, the above equation can be expressed in terms of an energy integral over $\eps_\vk$ using $\vf|\vk| = \vf\sqrt{2m(\eps_\vk + \mu)}$.

\begin{figure*}[h!btp]
    \includegraphics[width=\columnwidth]{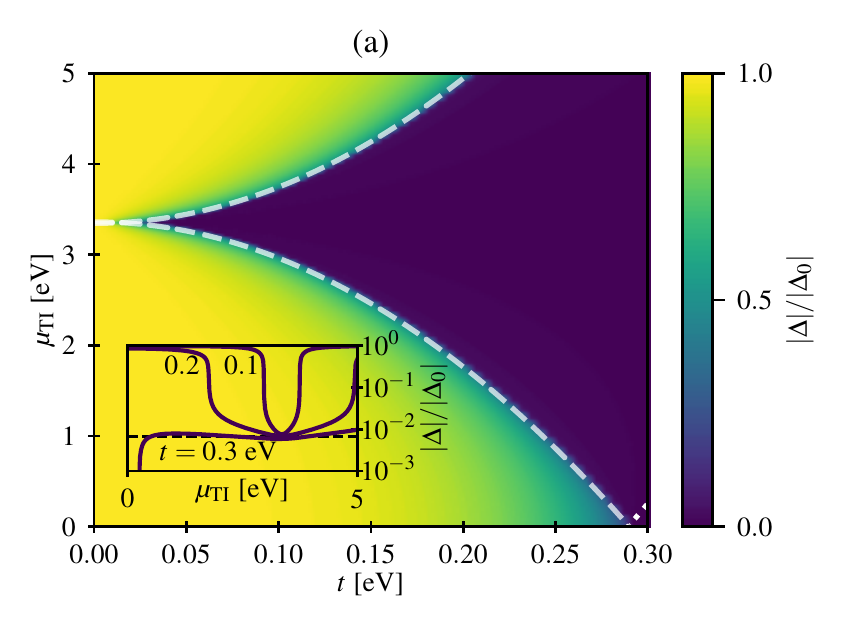}%
    \includegraphics[width=0.5\textwidth]{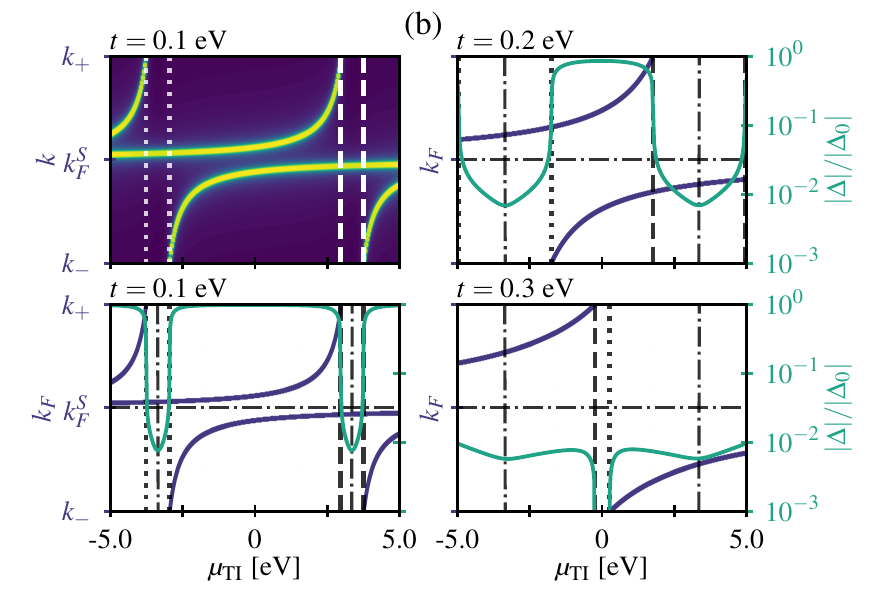}
    \caption{\label{fig:gap0} (a) Plot of the superconducting gap at $T=0$ for an $s$-wave superconductor as a function of $\muti$ and $t$ and  with an upper cutoff $\omega_{+} = 0.0025$ eV, normalized to the bulk value $|\Delta_0|$ for parameter values relevant for Nb-HgTe bilayers. The $k_F$-values for the $TI$ appear vertical on this plot as a function of $\mu_{TI}$ due to the small value of the cutoff  $\omega_{+}$. The numerical results show that the zero-temperature gap essentially is unaffected by the proximity to the TI for small values of $t$, where the suppression is severe only for values of $\muti$ close to $\sqrt{2m\vf^2\mu}$, a value far too large to be experimentally achievable. However, for increasing $t$, the region where superconductivity is suppressed increases quadratically with $t$, eventually leading to a suppression also for $\muti = 0$.
    {The inset shows the normalized gap at $t=\SI{0.1}{eV}$, $\SI{0.2}{eV}$ and $\SI{0.3}{eV}$, indicating that the gap is not suppressed entirely in most cases, {but rather to a reduced value of} $\Delta_0e^{-1/\lambda}$ (dashed line), consistent with there being only one band contributing to superconductivity in this region. The exception is close to $\muti=0$ for $t=\SI{0.3}{eV}$, where there are no bands with {Fermi wavevector between $k_-$ and $k_+$}, resulting in $\Delta=0$. This is the case in the area restricted by the dotted line in the main figure.}
    (b) The upper left panel is a plot of the integrand in the gap equation, Eq.~(\ref{eq:gapeq}) evaluated at $\Delta_0$ for {wavevectors $k_- < |\vk| < k_+$} and $t=\SI{0.1}{eV}$, where light colors correspond to high values of the integrand. The three remaining panels show the {magnitude of the Fermi wavevectors $k_F$} of the bands defined in Eq.~(\ref{eq:nonsc_bands}) (left axis) in the same interval at $t=\SI{0.1}{eV}$, $\SI{0.2}{eV}$ and $\SI{0.3}{eV}$, and the normalized gap (right axis). Notice that the plots are close to symmetric around $\muti=0$ since  $\omega_D \ll \mu$. The dash-dotted lines are {$k_F^S$ and $k_F^\mathrm{TI}(\muti)$, the Fermi wavevectors of} the S and TI for $t=0$, respectively.
    Comparing the two left panels it is clear that the main contribution to the integral in the gap equation comes from {wavevectors close to the Fermi wavevectors of the bands in the relevant $|\vk|$ interval.}
    $\muti^{\alpha, \pm}(t)$ are plotted as dashed ($\alpha=1$) and dotted ($\alpha=-1$) lines in all plots, indicating the onset of the region in parameter space where superconductivity is greatly suppressed.
    }
\end{figure*}

\section{Results and discussion}\label{sec:results}
From the expressions for the system eigenenergies in the non-superconducting case, Eq.~(\ref{eq:nonsc_bands}) we see that the S and TI bands have hybridized, leading to avoided crossings. The effect of this hybridization is largest when the chemical potential of both the S and TI is tuned such that the Fermi momenta coincide, i.e. for $\muti = \pm\sqrt{2m\vf^2\mu}$. {A possibly strong proximity effect should therefore be expected to occur in a region close to these values of $\muti$, the size of which increases with increased hopping $t$}. In the following we numerically solve the gap equations for both $s$- and $d$-wave superconductors for relevant parameter values.

\subsection{\texorpdfstring{$s$}{s}-wave pairing}

Using numerical values $\mu \sim \SI{5}{\electronvolt}$, a cut-off corresponding to the Debye frequency, $\hbar\omega_\pm = \hbar\omega_D\sim\SI{25}{\milli \electronvolt}$\cite{Ashcroft1976}, $\hbar^2/2m\sim \SI{40}{\milli \electronvolt \cdot \nano\meter \squared}$, $\hbar \vf \sim \SI{300}{\milli \electronvolt\cdot \nano\meter}$,\cite{Brune2011,Sochnikov2015} and $\lambda = 0.2$, we solve the gap equation in Eq.~(\ref{eq:gapeq}) for different values of $t$ and $\muti$ at $T=0$ for an $s$-wave superconductor.
The results in Fig.~\ref{fig:gap0}(a) show that the absolute value of the gap is not changed significantly due to the inverse proximity effect for small $t$, except for $\muti$ close to $\sqrt{2m\vf^2\mu}$. Both for $\muti$ above and below this region, the inverse proximity effect is small, signifying that the disappearing gap in the region where the inverse proximity effect is strong cannot be simply related to the increasing density of states in the TI. {For increasing $t$, the region where superconductivity is suppressed increases quadratically with $t$, eventually leading to suppressed superconductivity also at $\muti=0$.}


The strong suppression of the order parameter can be understood from the fact that the pairing potential is attractive only when $|\vk^2/2m -\mu|\le \omega_D$, {corresponding to wavevectors between $k_\pm \equiv \sqrt{2m(\mu \pm \omega_\pm)}$. This means that the Fermi wavevectors $k_F$ of the bands in Eq.~(\ref{eq:nonsc_bands}), the value of $|\vk|$ for which $\eps^\gamma_\alpha(\vk) = 0$, have to satisfy $k_-< k_F < k_+$} in order to contribute significantly to the integral in the gap equations and thus give a finite gap. This can be seen by comparing the left panels in Fig.~\ref{fig:gap0}(b), where the upper left panel shows the integrand of the gap equation, Eq.~(\ref{eq:gapeq}), and the lower left panel plots {$k_F$ for} the bands in Eq.~(\ref{eq:nonsc_bands}) as a function of $\muti$. The main contribution to the gap equation clearly comes from {the values $|\vk| = k_F$.}
From Fig.~\ref{fig:gap0}(b) we also see that as $\muti$ approaches $\pm\sqrt{2m\vf^2\mu}$, the value where the Fermi wavevectors for the bare the S and TI bands, {$k_F^S$ and $k_F^\mathrm{TI}(\muti)$} cross, the {wavevector} of one of the bands exceeds $k_+$ and thus does not contribute to the gap equation. Now there is only one non-degenerate band inside the relevant region, meaning that the density of states and thus $\lambda$ is halved compared to the $t=0$ case, where the band is doubly degenerate. Hence the resulting gap is suppressed to {$\Delta_0 e^{-1/\lambda} = 2\omega_De^{-2/\lambda}$}, in good agreement with the numerical results, as shown by the dashed line in the inset in Fig.~\ref{fig:gap0}(a). This also means that the suppression is less severe for higher $\lambda$, which we have confirmed by numerical simulations.

For positive $\muti$, {the Fermi wavevector} in one band exits the integration {interval $[k_-,k_+]$} at $\muti=\muti^{+,-}$, while a new band enters this region at $\muti=\muti^{+,+}$, where we have defined
\begin{align}
  \muti^{\alpha,\pm}(t) = \alpha\sqrt{2m\vf^2(\mu \mp \omega_D)} \pm \frac{t^2}{\omega_D}, \label{eq:muti_crit}
\end{align}
see appendix \ref{sec:proximity_criterion} for details. A similar argument holds for negative $\muti$, and hence superconductivity is strongly suppressed for
\begin{align}
  \muti^{\alpha,-} < \muti < \muti^{\alpha,+},\label{eq:prox_crit}
\end{align}
indicated by the dashed and dotted lines in Fig.~\ref{fig:gap0}. If the hopping parameter is large enough, $t^2 > \omega_D \sqrt{2m\vf^2(\mu\mp\omega_D)}\equiv (t_\mp)^2$, $\muti^{-,+}$ and $\muti^{+,-}$ change sign. Hence, for $|t|>|t_+|>|t_-|$ and $\muti^{+,-}<\muti<\muti^{-,+}$, no bands have a Fermi {wavevector between $k_-$ and $k_+$}, resulting in $\Delta=0$, as seen for $t\approx\SI{0.3}{eV}$ and low $\muti$ in Fig.~\ref{fig:gap0}. Since $\mu \gg \omega_D$, all results are close to symmetric about $\muti=0$, as seen in Fig.~\ref{fig:gap0}(b).

In order for strong suppression to occur for some value of $\muti$, we must require $\muti^{\alpha,-} < \muti^{\alpha,+}$. For $\alpha=-1$ this always holds, while for $\alpha = +1$ we get a lower limit for $t^2$,
\begin{align}
  t^2 > \omega_D\left[\sqrt{2m\vf^2(\mu+\omega_D)} - \sqrt{2m\vf^2(\mu-\omega_D)}\right].
\end{align}
For conventional $s$-wave superconductors $\mu \gg \omega_D$, meaning strong suppression can occur even at low values of $t$, though for TI chemical potentials close to $\pm\sqrt{2m\vf^2\mu}$.


While this result is strictly only valid in the limit of an atomically thin superconductor, we expect that this effect in principle could reduce the zero temperature gap and thus also reduce the critical temperature in superconducting thin films. {However, f}or typical parameter values in TIs and $s$-wave superconductors, the values of $\muti$ where superconductivity vanishes is inaccessible, tuning $\muti$ by several eV would place the Fermi level inside the bulk bands of the TI, where our model is no longer valid. The only exception from this is when $|t| \gtrsim |t_-|$, when superconductivity is suppressed even at $\muti=0$. The fact that no strong inverse proximity effect has been observed, e.g. in Ref.~\onlinecite{Sochnikov2015}, might indicate that the coupling constant $t$ is below this limit, meaning that an unphysical high chemical potential is needed in the TI to observe the vanishing of superconductivity. Since conventional $s$-wave superconductors have high Fermi energies, it might not be possible to reach the parameter regions where superconductivity vanishes, unless {the chemical potential in the S can be lowered significantly}, the Fermi velocity of the TI is {lowered by renormalization}, as was proposed in Ref.~\onlinecite{Sedlmayr2018}, or the coupling between the layers can be increased beyond $t_-$. However, {as we show below,} similar effects {are} present also for unconventional, high-$T_c$ superconductors, for which the Fermi energy is lower. Examples of such superconductors would be the high-$T_c$ cuprates and the heavy-fermion superconductors.\footnote{Although heavy-fermion superconductors nominally have a quite low critical temperature in absolute terms, they are nevertheless high-$T_c$ superconductors. Their critical temperatures are a significant fraction of their Fermi-temperatures.}

\begin{figure*}[h!btp]
    \includegraphics[width=\columnwidth]{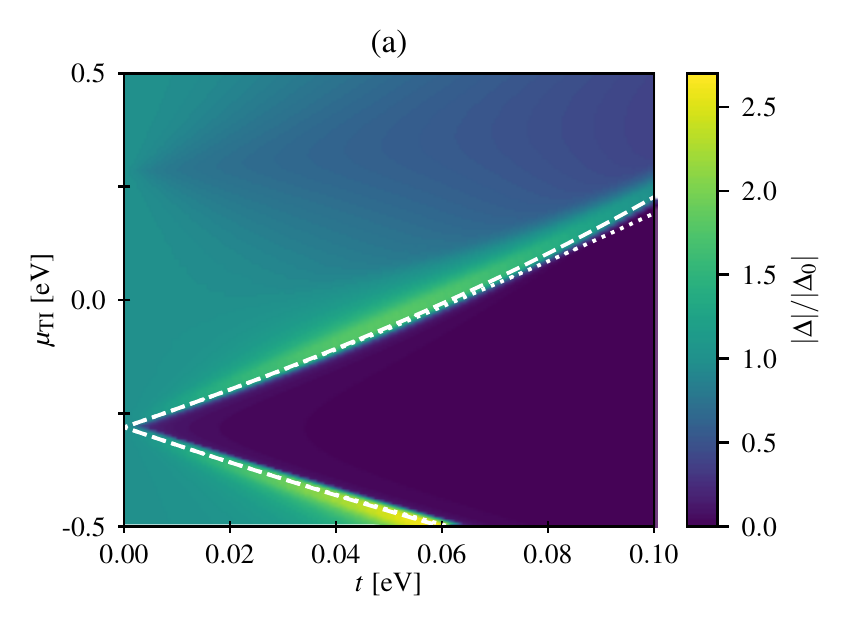}%
    \includegraphics[width=0.5\textwidth]{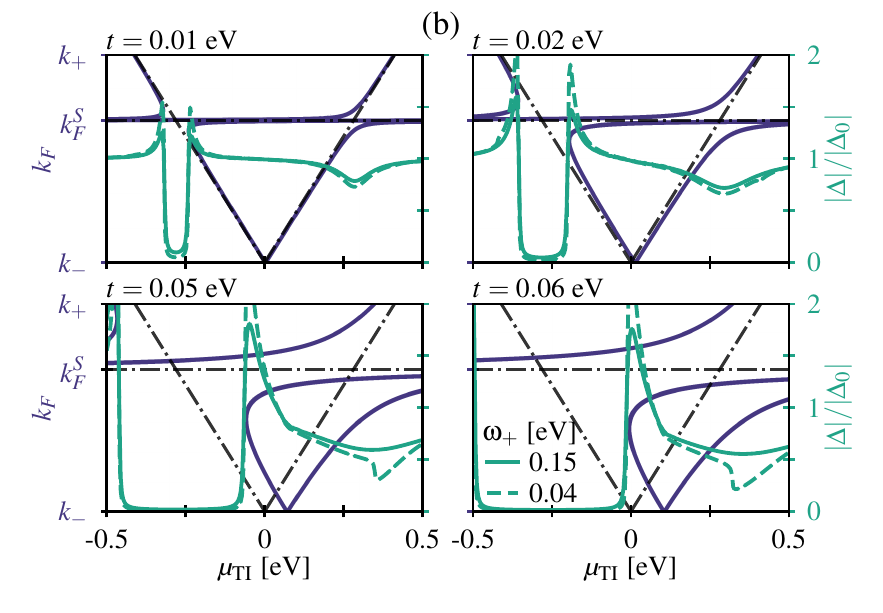}
    \caption{\label{fig:gap0_dwave} (a) Plot of the superconducting gap at $T=0$ for a $d$-wave superconductor as a function of $\muti$ and $t$ with upper cut-off $\omega_+ = \SI{0.15}{\electronvolt}$, normalized to the bulk value $|\Delta_0|$ for parameter values relevant for bilayers consisting of HgTe and high-$T_c$ superconductors. The gap is strongly suppressed for $\muti^- <\muti <\muti^+$, where the approximate (exact numerical) functions $\muti^\pm(t)$ in Eq.~(\ref{eq:mupm}) are plotted as dotted (dashed) lines. The approximate solution is only valid for {$k_F \approx k_F^S$}, corresponding to small $t$. For $\muti \approx \muti^\pm(t)$ the gap increases beyond $\Delta_0$.
    (b) Plot of the {magnitude of the Fermi wavevectors} of the bands in Eq.~(\ref{eq:nonsc_bands}) in the {interval $k_- < k_F < k_+$} (left axis), together with the normalized gap (right axis) for $\omega_+ = \SI{0.15}{eV}$ and $\SI{0.04}{eV}$. {The upper limit $k_+$ in the left axis corresponds to $\omega_+=\SI{0.04}{eV}$}. The black dash-dotted lines show the S and TI Fermi wavevectors for $t=0$. As for the $s$-wave case, the strong suppression of the gap is due to only one band having a Fermi {wavevector} in the integration interval. Note how the values of $k_F(\muti)$ of the hybridized bands (originating with the left $t=0$ crossing of the $k_F$s of the TI and the S) bend back in a pronounced was as a function of $\muti$ ($k_F$ is a multivalued function of $\muti$ since there are four bands). This leads to an enhanced density of states for these values of $\muti$. This in turn gives an  enhancement of the gap in the immediate vicinity of the region of $\muti$ where the gap is suppressed by the disappearance of bands crossing the TI Fermi surface. This effect is  not seen in the s-wave case, where the pronounced back bending of $k_F(\muti)$ does not occur {inside the integration interval} with the much lower values of $\omega_{\pm}$, see Fig.~\ref{fig:bands}(a). 
    }
\end{figure*}

\subsection{\texorpdfstring{$d$}{d}-wave pairing}

Using a much lower chemical potential in the S, $\mu \sim \SI{35}{\milli \electronvolt}$,\cite{Gerbstein1989} and an upper cut-off frequency comparable to the spin fluctuation energy in the high-$T_c$ cuprates, $\omega_+ \sim$ \SIrange{0.04}{0.15}{\electronvolt},\cite{Moriya1990,Monthoux1992,Nagaosa1997} $\omega_-=\mu$, and parameters otherwise as for the $s$-wave case, we solve the gap equations for a $d$-wave superconductor. First of all, the effect of the $d$-wave gap structure, compared to an $s$-wave gap, is an overall change in scaling, just as is the case for $\Delta_0$ (see Appendix \ref{sec:zerotempgap}). Hence, the results for $\Delta^{\textrm{$s$-wave}}/\Delta_0^{\textrm{$s$-wave}}$ are identical to $\Delta^{\textrm{$d$-wave}}/\Delta_0^{\textrm{$d$-wave}}$ when using the same parameters, and we have therefore solved the numerically more efficient $s$-wave gap equations with parameters valid for high-$T_c$ superconductors.

\begin{figure*}[h!tbp]
    \includegraphics[width=\textwidth]{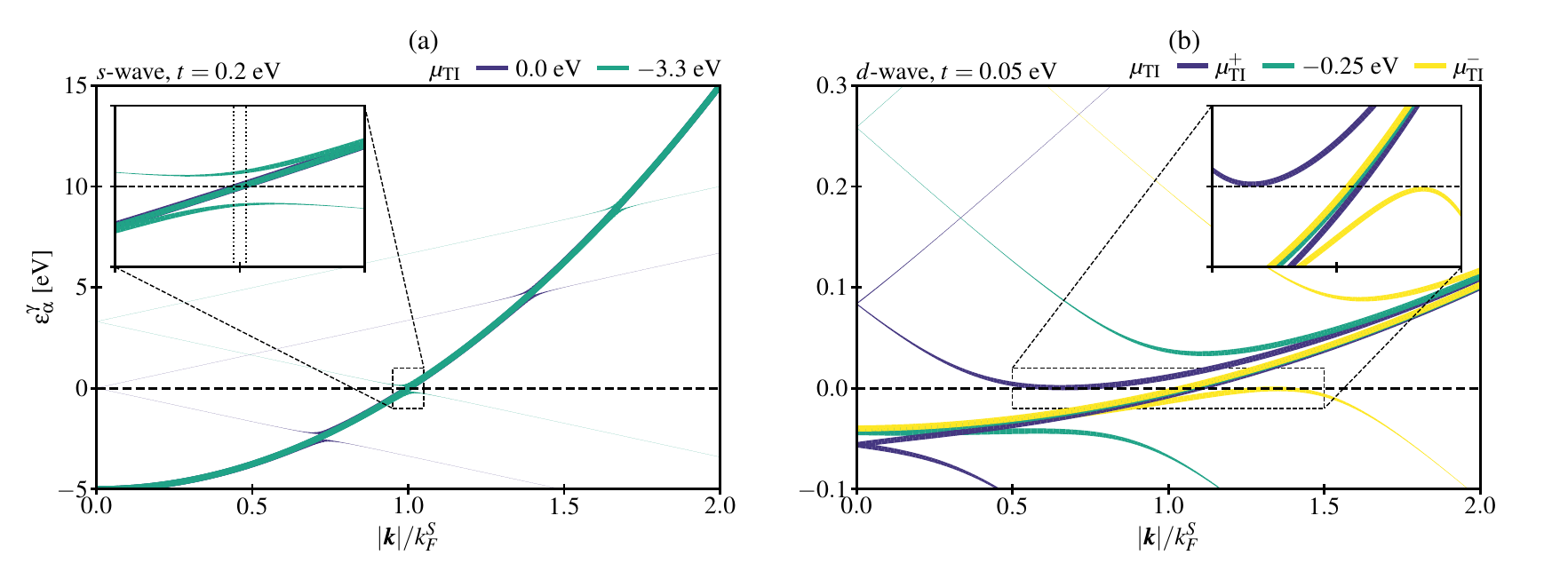}
    \caption{\label{fig:bands} Plots of the bands $\eps_\alpha^\gamma(\vk)$ in Eq.~(\ref{eq:nonsc_bands}) for (a) $s$-wave and (b) $d$-wave parameter values and different values of $\muti$. The line widths are proportional to the spectral weights $w_\alpha^\gamma(\vk)$ of the bands, see Eq.~(\ref{eq:residues}). In (a) the values of $\muti$ correspond to a barely suppressed ($\muti=\SI{0.0}{eV}$) and strongly suppressed ($\muti=\SI{-3.3}{eV}\approx -\sqrt{2m\vf^2\mu}$) gap for coupling $t=\SI{0.2}{eV}$. The inset shows that there is no hybridization of bands close to the Fermi level (dashed line) for the lowest $\muti$, while the strong hybridization for $\muti=\SI{3.3}{eV}$ leads to only one band crossing the Fermi level in the interval $[k_-, k_+]$ (dotted lines). In (b) we see that only one band crosses the Fermi level for $\muti = \SI{-0.25}{eV}$, explaining the strong suppression in this case. At $\muti=\muti^\pm$ we have an increase in $|\Delta|$, which can be explained by the bands having minima/maxima at the Fermi level in these cases, leading to high densities of states.} 
\end{figure*}

Fig.~\ref{fig:gap0_dwave}(a) shows the numerical results for the normalized gap as a function of $\muti$ and $t$. The most prominent difference compared to the results in Fig.~\ref{fig:gap0} is that the results are no longer symmetric about $\muti=0$, which can be understood from the fact that $\omega_\pm$ is of the same order of magnitude or larger than $\mu$. Due to the anti-crossing of the Fermi {wavevectors} at negative $\muti$, there is only one Fermi {wavevector between $k_-$ and $k_+$ for} $\muti^- < \muti < \muti^+$ (dashed lines in Fig.~\ref{fig:gap0_dwave}(a)), leading to strong suppression for negative $\muti$. This is illustrated in Fig.~\ref{fig:gap0_dwave}(b), where we plot the Fermi {wavevectors} of the bands together with the normalized gap as a function of $\muti$ for different values of $t$. The figure also shows how the regions of strong mixing between the bands increases with increasing $t$. Interestingly, the suppression of the gap is preceded by an increased $\Delta$ at $\muti^\pm$, due to the bending of the Fermi {wavevectors} away from the crossing point of {$k_F^S$ and $k_F^\mathrm{TI}(\muti)$},{ which leads to an increase in the density of states at the Fermi level. This is illustrated in Fig.~\ref{fig:bands}(b), where for TI chemical potentials $\muti^\pm$ the bands have a minimum (maximum) at the Fermi level, resulting in high densities of states. The difference in the gap enhancement between $\muti^+$ and $\muti^-$ is due to the combined effects of different spectral weights, indicated by the line widths in Fig.~\ref{fig:bands}(b), and the size of the Fermi surface, leading to a net larger increase in $|\Delta|$ at $\muti^-$.} In the small $t$ limit, we find the approximate expressions
\begin{align}
    \muti^\pm = - \sqrt{2m\vf^2\mu} \pm 2\left(\frac{m\vf^2}{2\mu}\right)^{1/4}t + \frac{1}{4\mu}t^2. \label{eq:mupm}
\end{align}
These lines are plotted in Fig.~\ref{fig:gap0_dwave}(a) (dotted lines) together with the exact numerical solutions (dashed lines), see Appendix \ref{sec:proximity_criterion} for details. {This increase in $|\Delta|$ is not due to the the $d$-wave symmetry, and should therefore be present for $\muti = \muti^\pm$ whenever the interval $[k_-, k_+]$ includes either of the points $k_F^S \pm |\delta k_F|$, where $\delta k_F$ is defined in Eq.~(\ref{eq:delta_k}).}

For positive $\muti$ there is a small reduction in $\Delta$ close to $\muti = \sqrt{2m\vf^2\mu}$, even though there are three bands with {$k_F \in [k_-, k_+]$}. However, since the numerator of each term in the gap equation Eq.~(\ref{eq:gapeq}) can be written $\xi^\pm_\alpha(\vk)^2 - (\alpha\vf|\vk|-\muti)^2$, regions where $\xi^\pm_\alpha(\vk)$ are similar to the bare TI bands contribute little to the gap equations, resulting in a small decrease of $\Delta$.

The effect of using a lower upper cut-off in the solution of the gap equations is also shown in Fig.~\ref{fig:gap0_dwave}. Comparing the $\omega_+ = \SI{0.15}{eV}$ and $\SI{0.04}{eV}$ lines, we see that for high $t$, the mixing of the S and TI bands is still significant at {$k_F=k_+$,} leading to abrupt changes in $\Delta$. For the negative $\muti$ the main effect of lowering the upper cut-off $\omega_+$ is a further increase of $\Delta$ at $\muti^\pm$.

\begin{figure}[h!tbp]
    \includegraphics[width=\columnwidth]{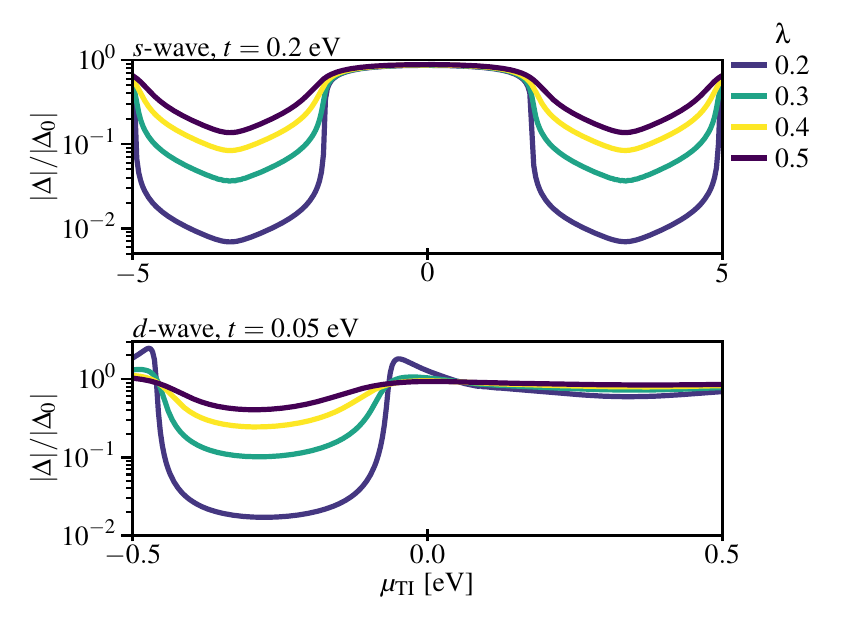}
    \caption{\label{fig:lambda_dep} The figure shows how the dimensionless coupling constant $\lambda$ affects the suppression of the superconducting gap for $s$-wave S with $t=\SI{0.2}{eV}$ (top) and $d$-wave S with $t=\SI{0.05}{eV}$ and $\omega_+ = \SI{0.15}{eV}$ (bottom). Increasing $\lambda$ makes the superconducting state more robust, reducing both the suppression of $\Delta$, and also the increase in $\Delta$ at $\muti^\pm$ in the $d$-wave case.}
\end{figure}

From the above results, it is clear that a strong suppression of the gap is more probable in S-TI bilayers consisting of a high-$T_c$ S, where both the chemical potential {$-\sqrt{2m\vf^2\mu}$ corresponding to $k_F^S = k_F^\mathrm{TI}(\muti)$}, and the hopping strength needed for strong suppression at $\muti=0$ is much lower. Hence, we may expect a strong inverse proximity effect in such systems, with a strength determined by $\lambda$, as illustrated in Fig.~\ref{fig:lambda_dep} for both the $s$- and $d$-wave case. Increasing $\lambda$ leads to a reduced suppression of the gap, consistent with the fact that the superconducting state is more robust for higher $\lambda$. {For the $s$-wave case, the suppression is proportional to $e^{-1/\lambda}$. This holds only approximately for the $d$-wave case due to other factors than Fermi level crossings affecting the suppression, such as changes in the spectral densities at the Fermi level and changes in the size of the Fermi surface (see Fig.~\ref{fig:bands}), effects which are small in the $s$-wave case.} From the results in Fig.~\ref{fig:gap0_dwave} we also see that it should be possible to change $\Delta$ by several orders of magnitude by small changes in $\muti$, again depending on the value of $\lambda$ as illustrated in Fig.~\ref{fig:lambda_dep}.

\section{Summary}\label{sec:summary}
We have theoretically studied the inverse superconducting proximity effect between a thin $s$-wave or $d$-wave superconductor and a topological insulator. Using a field-theoretical approach, we have found that in both cases there are regions in parameter space where the inverse proximity effect is strong, leading to a strong suppression of the gap approximately proportional to $e^{-1/\lambda}$. The suppression can be related to the hybridization of the TI and S bands, and the large degree of mixing which occurs when the Fermi wavevectors of the S and TI coincide for chemical potential $\muti=\pm\sqrt{2m\vf^2\mu}$. A larger value of $\lambda$ results in a more robust superconducting state, and hence less suppression.

For parameter values relevant for $s$-wave superconductors, the interval of suppression grows quadratically with the hopping $t$, and eventually leads to strong suppression even at $\muti = 0$. However, since there have been no experimental indications of a strong inverse proximity effect, we must conclude that the hopping is too weak to lead to suppression for experimentally accessible values of $\muti$. Neglecting the inverse proximity effect regarding the stability of the superconducting order therefore seems to be a good approximation for conventional $s$-wave superconductors.

A similar effect of suppressed superconductivity is also present for $d$-wave superconductors with parameter values relevant for the high-$T_c$ superconductors. In this case the strong suppression is found for TI chemical potentials close to $-\sqrt{2m\vf^2\mu}$, where the interval of strong suppression of the gap grows approximately linearly with $t$. Since the Fermi energy $\mu$ is much lower for high-$T_c$ superconductors, both the magnitude of the chemical potential $-\sqrt{2m\vf^2\mu}$, and the hopping strength needed for strong suppression at $\muti=0$ is much lower, making a strong inverse proximity effect more probable in such systems. In contrast to the $s$-wave case, the region of strong suppression was preceded by an increase in $\Delta$ above $\Delta_0$. {This is, however, not a consequence of the pairing symmetry, but rather the difference in system parameters. For large enough cut-off frequencies, the integration region will include a band minimum/maximum just touching the Fermi level, leading to a large increase in the density of states, and thus increased gap.}

We also find that the spin-triplet $p$-wave ($f$-wave) superconducting correlations are induced in the $s$-wave ($d$-wave) S due to the proximity-induced spin-orbit coupling.
Possible further work could include breaking the translation symmetry in the $x$ or $y$ direction and probing the density of states normal to the z-axis, possibly revealing signatures of $p$-wave or $f$-wave pairing. Moreover, it could be interesting to study the spatial variation of the order parameter in a superconductor with finite thickness.

\begin{acknowledgments}
J. L. and A. S.  acknowledge funding from the Research Council of Norway Center of Excellence Grant Number 262633, Center for Quantum Spintronics. A. S. and H. G. H. also acknowledge funding from  the Research Council of Norway  Grant Number 250985. J. L. acknowledges funding from Research Council of Norway Grant No. 240806. J. L. and
M. A. also acknowledge funding from the NV-faculty at the
Norwegian University of Science and Technology. H. G. H. thanks F. N. Krohg for useful discussions.
\end{acknowledgments}

\appendix

\section{Criteria for strong proximity effect}
\label{sec:proximity_criterion}

For superconductivity to occur, the Fermi wavevector of at least one of the bands has to lie within the interval of attractive pairing, which for $s$-wave superconductors is $\sqrt{2m(\mu-\omega_D)} < |\vk| < \sqrt{2m(\mu + \omega_D)}$.
We find the Fermi wavevector of the energy bands by setting $\eps_\alpha^\gamma(\vk) = 0$, which yields the equation
\begin{align}
    \left[\alpha\vf|\vk|-\muti\right]\eps_\vk - t^2 = 0.\label{eq:fermi_lvl_condition}
\end{align}
Inserting $|\vk| = k_ \pm$ we get the value of $\muti$ for which the Fermi wavevector of a band enters or leaves the interval of attractive pairing,
\begin{align}
  \muti^{\alpha,\pm}(t) = \alpha\sqrt{2m\vf^2(\mu \mp \omega_D)} \pm \frac{t^2}{\omega_D}.
\end{align}
The Fermi wavevectors of the bands $\eps_\alpha^-(\vk)$ exceed $k_+$ at $\muti^{\alpha,-}$, while the Fermi wavevectors of $\eps_\alpha^+(\vk)$ enter the interval $[k_-, k_+]$ at $\muti^{\alpha,+}$.
$\muti^{+, +}$ ($\muti^{-, -}$) is always positive (negative), while $\muti^{+, -}$ and $\muti^{-, +}$ change sign when $t^2 > \omega_D\sqrt{2m\vf^2(\mu +\omega_D)}\equiv (t_0^+)^2$ and $t^2> \omega_D \sqrt{2m\vf^2(\mu-\omega_D)}\equiv (t_0^-)^2$ respectively, where $|t_0^+| > |t_0^-|$.

Hence we have strong suppression when
\begin{align}
    \muti^{\alpha,-}<\muti<\muti^{\alpha,+},
\end{align}
which for $\alpha=+1$ requires $$t^2 > \omega\left[\sqrt{2m\vf^2(\mu+\omega_D)} - \sqrt{2m\vf^2(\mu-\omega)}\right].$$ Moreover, for $|t|>|t_+|$ and $\muti^{+,-}<\muti<\muti^{-,+}$ no bands have a Fermi {wavevector} inside the relevant interval, and the gap is zero.




For the $d$-wave S we find an increase in the gap function for certain values of $\muti$. An increase in the gap would occur in regions where the Fermi wavevectors of two bands approach each other and finally coincide as a function of $\muti$, resulting in a {region of closely spaced Fermi wavevectors}. This can be seen to happen in Fig.~\ref{fig:gap0_dwave}(b). To find the value of $\muti$ corresponding to the increase in $\Delta$ we find the local minima of {
\begin{align}
    \muti(k_F) = \alpha\vf k_F - \frac{t^2}{\eps_{k_F}}\label{eq:muti_eF}
\end{align}
by requiring $\del_{k_F} \muti(k_F) = 0$, from which we get the equation for ${k_F}$
\begin{align}
    \alpha \vf + \frac{t^2 k_F}{m \eps_{k_F}^2} = 0.
\end{align}
Solving this equation numerically with $\alpha=-1$ and inserting the results into Eq.~(\ref{eq:muti_eF}) yields the dashed lines in Fig.~\ref{fig:gap0_dwave}, in good agreement with the numerical results of the gap equation.
To get an approximate analytical expression, we assume that $k_F = k_F^S + \delta k_F$, where $\delta k_F \ll k_F^S$, which is valid for sufficiently small $t$. Neglecting terms of $\mathcal{O}(\delta k_F^3)$ and higher, we get
\begin{align}
    \delta k_F^2 + \frac{t^2m}{\alpha \vf k_F^S} + \frac{t^2m}{\alpha \vf (k_F^S)^2}\delta k_F = 0.
\end{align}
Neglecting the last term yields, effectively keeping terms up to $\mathcal{O}(t^2)$, results in
\begin{align}
    \delta k_F = {}& \pm \sqrt{-\frac{1}{\alpha}}\left(\frac{m}{2\vf^2\mu}\right)^{1/4}t,\label{eq:delta_k}
\end{align}
from which it is clear that we only have solutions for $\alpha=-1$. Inserting this expression into Eq.~(\ref{eq:muti_eF}), we get to $\mathcal{O}(t^2)$
\begin{align}
    \muti^\pm \approx - \sqrt{2m\vf^2\mu} \pm 2\left(\frac{m\vf^2}{2\mu}\right)^{1/4}t + \frac{1}{4\mu}t^2.
\end{align}
} This result is plotted as dotted lines in Fig.~\ref{fig:gap0_dwave}(a), and is in good agreement with the exact numerical results for small $t$. For $\muti^- < \muti < \muti^+$, there is only one Fermi {wavevector} in the integration region, leading to a suppressed gap.

\section{Functional integral in Nambu spinor notation}
\label{sec:Nambu_funcint}
We begin by considering the Gaussian integral over Grassmann variables,\cite{Wegner2016}
\begin{align}
    I &= \left(\prod_i\int \rmd a_i\right) ~ e^{-\frac{1}{2}\sum_{i,j} a_iM_{ij}a_j} \nonumber\\
    &= \left(\prod_i\int\rmd a_i\right)\prod_{i,j}(1 - \frac{1}{2}a_iM_{ij}a_j) = \operatorname{Pf}\left(\frac{M-M^T}{2}\right)\label{eq:Pfaffian},
\end{align}
where $\operatorname{Pf}((M-M^T)/2)$ is the Pfaffian of the antisymmetric part of $M$, where $\operatorname{Pf}(A)^2 = \det(A)$.
As an example we consider only two variables, $a_1$ and $a_2$. In this case, terms containing $M_{ii}$ disappear, since $a_i^2 = 0$, as do second order terms in $M$. For the above integral we therefore get
\begin{align}
    I &= \int \rmd a_1 \rmd a_2 ~ \frac{1}{2}(-a_1 M_{12} a_2 - a_2 M_{21} a_1) = \frac{M_{12} - M_{21}}{2} \nonumber\\
    &= \sqrt{\det \frac{M - M^T}{2}} = \sqrt{\det M^A} = \operatorname{Pf}(M^A).
\end{align}
Here, $M^A$ is the anti-symmetric part of $M$.


Applying this to the problem of integrating $\exp(-S_\rm{S}^\rm{eff})$, we first write the action in terms of the Nambu spinor $\C$:
\begin{widetext}
\begin{align}
    S_\rm{S}^\rm{eff} &= -\frac{1}{\beta V}\sum_{k,k'} \C^T(-k) \begin{pmatrix}
    \varphi^\dagger(k'-k)\frac{\s_x-i\s_y}{2} & 0\\
    \G_0^{-1}(k)\delta_{k,k'} & \varphi(k-k')\frac{\s_x+i\s_y}{2}
    \end{pmatrix}\C(k') \equiv -\frac{1}{2\beta V}\sum_{k,k'} \C^T(-k)  A(k,k') \C(k') \nonumber\\
    &=-\frac{1}{\beta V}\sum_{k,k'} \C^T(k) \begin{pmatrix}
    -\varphi^\dagger(k-k')\frac{\s_x+i\s_y}{2} & -[\G_0^{-1}(k)]^T\delta_{k,k'}\\
    0 & -\varphi(k'-k)\frac{\s_x-i\s_y}{2}
    \end{pmatrix}\C(-k')\equiv -\frac{1}{2\beta V}\sum_{k,k'} \C^T(k)  [-A(k',k)]^T \C(-k').\nonumber
\end{align}
Combining these two expressions, we get
\begin{align}
    S_\rm{S}^\rm{eff} &= -\frac{1}{2\beta V}\sum_{k,k'} \C^T(-k) \begin{pmatrix}
    -\varphi^\dagger(k'-k)i\s_y & -[\G_0^{-1}(-k)]^T\delta_{k,k'}\\
    \G_0^{-1}(k)\delta_{k,k'} & \varphi(k-k')i\s_y
    \end{pmatrix}\C(k') \nonumber\\
    &= -\frac{1}{2\beta V} \sum_{k,k'} \C^T(-k) \frac{A(k,k') - A^T(-k',-k)}{2} \C(k') = -\frac{1}{2\beta V} \sum_{k,k'} \C^T(-k) A^A(k,k') \C(k'),\label{eq:Seff_CT}
\end{align}
where $A^A(k,k')$ denotes the anti-symmetric part of $A$. {This is exactly equal to Eq.~(\ref{eq:Seff_nambu}), as can be seen by the following manipulations. For notational simplicity we use the $2$-vector notation}
\begin{equation}\C(k) = \begin{pmatrix}
    c(k)\\
    c^*(-k)
\end{pmatrix},
\end{equation}
{i.e $[\C(k)]_1 = c(k),~[\C(k)]_2 = c^*(-k)$. Hence the matrix multiplication in Eq.~(\ref{eq:Seff_CT}) can be written}
\begin{align}
   \sum_{ij} [\C^T(-k)]_i [A^A(k,k')]_{ij} [\C(k')]_j =
   {}&-[\C^T(-k)]_1 \varphi^\dagger(k'-k)i\s_y [\C(k')]_1
   - [\C^T(-k)]_1 [\G_0^{-1}(-k)\delta_{k,k'}]^T [\C(k')]_2\nonumber\\
   &+[\C^T(-k)]_2 \G_0^{-1}(-k)\delta_{k,k'} [\C(k')]_1
   +[\C^T(-k)]_2 \varphi(k-k')i\s_y [\C(k')]_2.
\end{align}
{We use the fact that $[\C^\dagger(k)]_1 = [\C^T(-k)]_2$ and $[\C^\dagger(k)]_2 = [\C^T(-k)]_1$, and relate the remaining factors to the elements of $\G^{-1}(k,k')$ in Eq.~(\ref{eq:Ginv}),}
\begin{align}
   C^T(-k) A^A(k,k') C(k') = {}&[\C^\dagger(k)]_2[\G^{-1}(k,k')]_{21}[\C(k')]_1 + [\C^\dagger(k)]_2[\G^{-1}(k,k')]_{22}[\C(k')]_2 \nonumber\\
   &+[\C^\dagger(k)]_1[\G^{-1}(k,k')]_{11}[\C(k')]_1 + [\C^\dagger(k)]_1[\G^{-1}(k,k')]_{12}[\C(k')]_2\nonumber\\
 ={} & \C^\dagger(k) \G^{-1}(k,k')\C(k'),
\end{align}
\end{widetext}
which shows that Eq.~(\ref{eq:Seff_CT}) and Eq.~(\ref{eq:Seff_nambu}) are equivalent.
Using Eq.~(\ref{eq:Pfaffian}), the functional integral of the action in Eq.~(\ref{eq:Seff_CT}) results in
\begin{equation}
    Z = \int \D c^\dagger \D c ~ e^{-S_\rm{S}^\rm{eff}} = \sqrt{\det\left[-A^A\right]},
\end{equation}
where we have neglected various numerical constants. By interchanging an even number of rows, it can be shown that $A^A(k,k') \rightarrow \G^{-1}(k,k')$, and since the determinant is invariant under an even number interchanges, we find\cite{Krohg2018}
\begin{equation}
    Z = e^{\frac{1}{2}\Tr \ln (-\G^{-1})}.
\end{equation}

\section{Zero temperature gap for \texorpdfstring{$t=0$}{t=0}}
\label{sec:zerotempgap}
When $t=0$, the gap equation, Eq.~(\ref{eq:gapeq}), reduces to
\begin{align}
  1 = \frac{g}{2V}\sum_\vk \frac{v^2(\vk)}{\sqrt{\eps_\vk + |\Delta_0(\vk)|^2}}
\end{align}
in the zero temperature limit. Transforming this to an integration over $\phi_\vk$ and energy, we get
\begin{align}
  1 &= \frac{\lambda}{2} \int_{-\omega_-}^{\omega_+}\rmd \eps \int_0^{2\pi}\frac{\rmd \phi_\vk}{2\pi} \frac{v^2(\phi_\vk)}{\sqrt{\eps + |\Delta_0(\phi_\vk)|^2}},
\end{align}
where $\omega_\pm$ are positive. Performing the energy integral we get
\begin{align}
1  &= \frac{\lambda}{2} \int_0^{2\pi}\frac{\rmd \phi_\vk}{2\pi}v^2(\phi_\vk) \ln\frac{\sqrt{|\Delta_0(\phi_\vk)|^2 + \omega_+^2} + \omega_+}{\sqrt{|\Delta_0(\phi_\vk)|^2 + \omega_-^2}-\omega_-^2}\nonumber\\
  &\approx \frac{\lambda}{2} \int_0^{2\pi}\frac{\rmd \phi_\vk}{2\pi}v^2(\phi_\vk) \left[\ln\frac{4\omega_-\omega_+}{\Delta_0^2} - 2\ln|v(\phi_\vk)|\right],
\end{align}
where we in the last line have assumed that the gap is small compared to the cut-off energy. For an $s$-wave superconductor $v(\phi_\vk) = 1$, and we get simply $\Delta_0 = 2\sqrt{\omega_-\omega_+} e^{-1/\lambda}$. For $d$-wave pairing we can instead write the gap as
\begin{align}
  \Delta_0 = 2\sqrt{\omega_-\omega_+} e^{-\frac{1}{\lambda} - I},
\end{align}
where we have defined the integral
\begin{align}
  I = \int_0^{2\pi} \frac{\rmd \phi_\vk}{2\pi} v^2(\phi_\vk)\ln|v(\phi_\vk)| = \frac{1-\ln2}{2} \approx 0.153426.
\end{align}
Hence, the maximum $d$-wave gap-amplitude is marginally smaller than the $s$-wave gap for the same values of $\lambda$ and $\omega_\pm$.

\section{Numerical integration procedures}
When solving the gap equation numerically, the $\vk$ sum is rewritten in terms of an energy integral over $\eps_\vk$ and an integral over $\phi_\vk$, which in the $s$-wave case is simply equal to $2\pi$. In the $s$-wave case we therefore only have to perform the energy integral for energies in the interval $[-\omega_-, \omega_+]$, in our case using Python and the implementation \verb|trapz| of the trapezoidal method in the \verb|scipy| library. In the $d$-wave case, we use the \verb|quadpy| library's implementation of the numerical integration method in Ref.~\onlinecite{Xiao2010} when calculating the 2D integral in the $\eps_\vk- \phi_\vk$ plane.

\end{document}